\documentclass[bib]{ba}

\usepackage{bayes}
\usepackage[utf8]{inputenc}

\usepackage{amsmath,amsfonts,amsthm,amssymb,latexsym}

\usepackage{natbib}

\newcommand\TAP{{\em Théorie Analytique des Probabilités}}
\usepackage{ tipa }
\renewcommand\th{\textOlyoghlig}

\newenvironment{myquote}
   {\smallskip\begin{quote}\small\begin{em}}
   {\normalsize\end{em}\end{quote}\smallskip}
\newcommand\beq{\begin{myquote}}
\newcommand\enq{\end{myquote}}

\begin{document}

\inserttype{article}
\author{C.P.Robert}{Christian P.~Robert\\Universit\'e Paris-Dauphine, CEREMADE, IUF, and CREST, Paris\\
  {\sf xian@ceremade.dauphine.fr}}

\title[A first read of Théorie Analytique des Probabilités]{Reading {\em Théorie Analytique des Probabilités}}

\maketitle

\begin{abstract}
This note is an extended read of my read of Laplace's book \TAP, 
when considered from a Bayesian viewpoint but without
historical nor comparative pretentions. A deeper analysis is provided in Dale (1999).
\end{abstract}

\section{Introduction}
\label{sec:intro}

    \beq ``The theory of probabilities draws a remarkable distinction between observations which have been made, and
those which are to be made." {\em A. de Morgan, {\em Dublin Review}, 1837.}\enq

Pierre Simon Laplace's book, \TAP, was first published in \citeyear{laplace:1812}, that is, exactly two
centuries ago! Following a suggestion by the editor of the ISBrA Bulletin, I gladly accepted the invitation as
(a) Laplace's role in Bayesisian statistics is much deeper and longlasting than Bayes'
\citep{dale:1982,dale:1999}, (b) I had never looked at this book and so this was a perfect opportunity to do
so, using the 1812 edition in my possession, and (c) I was curious to see how much of the book had permeated
modern probability and statistics. (Note that the versions of the book evolved quite considerably from the
first to the fifth edition in 1825.) The following review is not pretending at scholarly grounding the book
within its academic surroundings and successors, but is to be taken as a mere Bayesian excursion along its
pages. A deeper analysis of \TAP~can be found in \citeauthor{dale:1999} (1999, pp.~250--283). In particular,
Andrew Dale discusses Bayesianly relevant supplements found in later editions of \TAP, as well as connections
with both Bayes' and Laplace's {\em Essays}.

    \beq ``Je m'attache surtout, à déterminer la probabilité des causes et des résultats indiqués par événemens
considérés en grand nombre." {\em P.S. Laplace, \TAP, page 3.}\enq

I must first and foremost acknowledge I found the book rather difficult to read and this for several reasons:
(a) as always is the case for older books, the ratio {\em text-to-formulae} is very high; (b) the themes in
succession are often abruptly brought (i.e.~not always well-motivated) and uncorrelated with the previous ones;
(c) the mathematical notations are (unsurprisingly) 18th-century, so sums are indicated by $S$, exponentials by
$c$, and so on, while those symbols are also used as variables in other formulae; (d) I often missed the big
picture and got mired into technical details, until they made sense or until I gave up; (e) I never understood
whether or not Laplace was interested in the analytics like generating functions only to provide precise
numerical approximations or for their own sake. So a certain degree of disappointment in the end, most likely due to my
insufficient investment in the project (on which I only spent an Amsterdam/Calgary flight and a few sleepless
nights in Banff...), even though I got excited by finding the bits and pieces about Bayesian estimation and
testing.

\section{Contents of \TAP}

    \beq ``Sa théorie est une des choses les plus curieuses et les plus utiles que l'on ait trouvées sur les suites."
{\em P.S. Laplace, \TAP, page 8.}\enq

The {\em Livre Premier} is about generating functions (Calcul des Fonctions géneratrices). As such, it is not
directly of interest, focusing on finite difference equations, even though the techniques developped therein
will be exploited in the second part. (There is an interesting connection with Abrahama de Moivre,
incidentally, since this older mathematical giant used generating functions to derive binomial formulas. He is
acknowledged in Laplace's preface by the above quote, \citealp{bellhouse:2011}.)

    \beq ``La théorie des probabilités consiste à réduire tous les événemens qui peuvent avoir lieu dans une
circonstance donnée à un certain nombre de cas également possibles." {\em P.S. Laplace, \TAP,
page 178.}\enq

The {\em Livre Second} is about probability theory, first about urn type problems, then about asymptotic
approximations. The introduction to this second part reflects the famous (almost mythical!) determinism of
Laplace, where randomness is simply l'expression de notre ignorance (yes, our ignorance as so expressed, page
177)... The intial pages contain the basics of probability like the chain rule, the product rule, the
conditional probability and what we now call Bayes' rule, even though it is not called as such in \TAP. I did
not find any mention of Thomas Bayes in the book. However, when looking at the on-line version of the book, I
realised to my dismay that the 1814 edition has changed quite significantly, with an historical introduction to
the theory of probability, incl.~the mention of Bayes. (Thus, the changes were not restricted to the removal of
the dedication to Napoléon-le-Grand [not longer appropriate after Waterloo and the restauration of the
monarchy!] and the change from Chancellier du Sénat [an honorific title under Napol\'eon Ier] to Pair du
Royaume [an honorific title under Louis XVIII], reflecting the well-known turncoat politics of Laplace!) An
interesting syntactic point is the paragraph where Laplace introduces the notion of {\em expectation} (in the
sense of Dicken's {\em Great Expectations}), along with {\em fears} ("crainte"), and as in Laplace's {\em Essai
philosophique}, he distinguishes between mathematical expectation and moral expectation. (He later acknowledge
Bernoulli’s priority, as discussed below.)

    \beq ``Nous traiterons d'abord les questions dans lesquelles les probabilités des événemens simples, sont
données; nous considérerons ensuite celles dans lesquelles ces probabilités sont inconnues, et doivent être
déterminées par les événemens observés." {\em P.S. Laplace, \TAP, page 188.}\enq

The above quote is the introduction to Chapter II which essentially consists in a sequence of combinatorial
problems solved by polynomial decompositions and approximated by the finite difference formulae of the first
{\em Livre}. (Despite this enticing quote, the chapter does not cover the statistical part.) While the accumulation of
lottery and urn problems is not exactly fascinating, to say the least, some entries highlight Laplace's analytical skills. For
instance, a convoluted urn problem leads to an equally convoluted integral (page 222)
$$
\dfrac{\int_0^\infty x^{rn-n} dx \cdot (x-r)^n e^{-x}}{\int_0^\infty x^{rn-n} dx \cdot e^{-x}} \qquad (0)
$$
where Laplace uses a Laplace approximation  to replace (0) with
$$
\dfrac{(1-1/n)^{n+1}}{\sqrt{(1-1/n)^2+\frac{2}{rn}-\frac{1}{rn^2}}}
$$
for $N$ and $rn$ large. The cdf is used in a convoluted (if labeled as ``très-simple" on page 264!) derivation
of an expectation of several variables. The chapter concludes with reflections on an optimal voting system that
relates to Condorcet's (although no mention is made of this political scientist in the book, even though
Laplace owed his position [at the age of 24!] in the Acad\'emie Royale des Sciences to his intervention).

    \beq ``On peut encore, par l'analyse des probabilités, vérifier l'existence ou l'influence de certaines causes
dont on a cru remarquer l'action sur les êtres organisés." {\em P.S. Laplace, Théorie Analytique des Probabilités,
page 358.}\enq

Chapter III moves to asymptotic approximations and the law of large numbers for frequencies, ``cet important
théorème" (page 275). The beginning of the chapter shows that the variation of the empirical frequency around
the corresponding probability is of order $1/ \surd n$, with a normal approximation to the coverage of the
confidence interval. \cite{dale:1999} makes the crucial point (and I missed it!) that Laplace defines there a
confidence interval on a probability parameter $p$, by a Bayesian argument, i.e. by using a flat prior on the
probability parameter (page 254).

    \beq ``On peut reconnaître l'effet très-petit d'une cause constante, par une longue suite d'observations dont les
erreurs peuvent excéder cette effet lui-même." {\em P.S. Laplace, \TAP, page 352.}\enq

Chapter IV extends the above law of large numbers to a sum of iid variables. It then remarks that the most
likely error is zero (which simply means that the mode of the standard normal distribution is indeed zero). It
also contains a derivation of (a) the posterior median as minimising the absolute error loss and
(b) the empirical average as minimising the squared error error or being the least square estimator
(page 321).  I think Laplace uses a Fourier transform to derive the distribution of a weighted sum (page 314).
Laplace then proceeds to generalise this optimality result to a bivariate quantity, obtaining again the least
square estimate and computing a bivariate Gaussian density on the way. And then comes the major step,!
namely Laplace's derivation of a posterior distribution (page 334):
$$
\dfrac{\prod_i \varphi(x_i-\theta)}{\int\prod_i \varphi(x_i-\theta)\,\text{d}\theta}
$$
(with my notations), thus using a flat prior on the location parameter! This fundamental step is compounded by
the introduction of a (not yet) Bayes estimator minimising posterior absolute error loss and found to be the
median of the posterior. In the next pages, Laplace attempts to find the MAP (which is also the maximum
likelihood estimator in this case), as an approximation to the posterior median (page 336). From therein, he
moves to identify the distribution for which the MAP is also the (arithmetic) average, ending up with the
normal distribution (page 338). (This result was to be extended by J.M. Keynes, see \citealp{keynes:1921}, to
different types of estimators.) The chapter concludes with a defense of the arithmetic mean as a limiting Bayes
estimator that does not depend on the law of the errors.

    \beq ``Pour déterminer avec quelle probabilité cette cause est indiquée, concevons que cette cause n'existe
point." {\em P.S. Laplace, \TAP, page 350.}\enq

Chapter V starts with the computation of a p-value, nothing less! Laplace analyses the likelihood
({\em vraisemblance}) of a non-zero effect by looking at the cdf of the observation under the null (page 361). The
following pages discuss Laplace's analysis of the irregularities in celestial trajectories, like the
perturbations between Saturn and Jupiter. It argues in a philosophical if un-Popperian way about the importance
of probabilistic analysis (read statistics) for uncovering scientific facts (page 358).

    \beq ``Laplace actually used the theory of probabilities as a method of discovery." {\em A. de Morgan, {\em
Dublin Review}, 1837.}\enq

In Chapter VI, {\em De la probabilité des causes et des événemens futurs, tirés des événemens observés}, Laplace
develops his Bayesian (or Laplacian) perspective for drawing inference about unknown probabilities. He uses a
uniform prior (with an interesting argument transferring the prior into the likelihood as to always consider
this case, see page 364).\footnote{As pointed out by Jean-Louis Foulley (personnal communication), this idea of
representing the non-uniform prior as an additional set of data independent of the observation is very
innovative. In modern Bayesian statistics language, t leads to easy and useful interpretations for conjugate
priors and may even be viewed as the basic idea behind partial (intrinsic and fractional) Bayes Factors.}  He
then derives a normal approximation to the posterior (first term of the Laplace approximation!, page 367). This
chapter also contains the famous study on the proportion $\varrho$ of female births in Paris, using an
approximation to the beta integral to show that the (posterior) probability that $\th$ is larger than 1/2 is
negligible (``d'une petitesse excessive", page 380).  Laplace also computes the posterior probability that the
probability of a male birth in London is larger than in Paris, which he finds equal to 1-1/328269 (using a
double integral and a continued fraction approximation!). He then moves to the applications of these techniques
to mortality tables and insurances, exhibiting there a thematic connection \citep{bellhouse:2011} with Abraham
de Moivre (and maybe even Bayes!). The chapter concludes by a computation of the posterior (or predictive!)
probability that $1-\varrho$ will remain larger than 1/2 in the next century, obtaining  a value of $0.782$.
							
Chapter VII is a short chapter on biased coins and compounded experiments, not directly related with Bayesian
perspectives (\citealp{dale:1999} extrapolates on this point, since the imprecision on the coin biasedness can
be seen as a prior). Chapter VIII is similarly short, reproducing earlier normal approximations on averages of
life durations. It also contains an interesting study on the impact of removing the impact of smallpox on the
death rate. Chapter IX deals with expectations of simple functions for binomial experiments and with their
normal approximation, again exhibiting the above link with de Moivre's on life insurrances.

Chapter X returns to the notion of moral expectation mentioned both earlier and in Laplace's {\em Essai
Philosophique}. The core (to solving the Saint Petersburg paradox) is to use $\log(x)$ instead of $x$ as a
utility function, following Bernoulli's derivation (now mentioned on page 439).

\section{Reflections}

    \beq ``In reviewing the general design of the work of Laplace, we desire to make the description of a book mark
the present state of a science." {\em A. de Morgan, {\em Dublin Review}, 1837.}\enq

In conclusion, {\em Théorie Analytique des Probabilités} provides a fascinating historical perspective on
Laplace's genius in framing probability and statistics within mathematical analysis and in deriving numerical
approximations to intractable integrals. As put by Augustus de Morgan in a praising if sometimes hilarious
review of the book, ``Théorie des Probabilités is the Mont Blanc of mathematical analysis". (Morgan considers
that the French national school of mathematics neglects to credit predecessors. It is quite true that it is
impossible to gather which results are original and which are not in  Théorie Analytique des Probabilités. He
similarly thinks that the first part on generating functions is mostly useless for the second part. And that
the introduction [in the 1814 edition] is the {\em Essai Philosophique}, whose final version is much enlarged
compared with this introduction.  Interestingly, de Morgan also spends quite some time on the notion of moral
expectation.) As opposed to Thomas Bayes' \citeyear{bayes:1763} short essay,\footnote{\cite{dale:1999} compares
Bayes' and Laplace's input, making the significant remark that Bayes considers ``a single urn" while "Laplace
entertained the idea of a {\em population} of urns" (p.277). This is a very powerful distinction, in that it
highlights how closer Laplace was from the notion of prior distribution.} the book by Laplace leads to a
global vision of the role and practice of probability theory, as it was then understood at the beginning of the
19th Century, and it can be argued the Théorie Analytique des Probabilités shaped the field (or fields) for
close to a hundred years.\footnote{It thus came as a surprise to read that Laplace was so much scorned and
despised by the statisticians of the mid-1800's and even far into the 1900's, see \cite{mcgrayne:2011}.}

\section*{ACKNOWLEDGEMENTS}
This research is supported partly by the Agence Nationale de la Recherche through the 2009-2012 grants {\sf Big
MC} and {\sf EMILE} and partly by the Institut Universitaire de France (IUF). It was undertaken during the BIRS
12w5105 meeting on ``Challenges and Advances in High Dimensional and High Complexity Monte Carlo Computation
and Theory", thanks to the superb conditions at the Banff Centre. The author is also grateful to
Julien Cornebise for providing him with a 1967 {\em fac-simile} reproduction of the 1812 edition. Comments from
Jean-Foulley were quite helpful in preparing the final version of this review.


\begin{thebibliography}{7}
\expandafter\ifx\csname natexlab\endcsname\relax\def\natexlab#1{#1}\fi
\expandafter\ifx\csname url\endcsname\relax
  \def\url#1{\texttt{#1}}\fi
\expandafter\ifx\csname urlprefix\endcsname\relax\def\urlprefix{URL }\fi
\providecommand{\eprint}[2][]{\url{#2}}

\bibitem[{Bayes(1763)}]{bayes:1763}
\textsc{Bayes, T.} (1763).
\newblock An essay toward solving a problem in the doctrine of chances.
\newblock \textit{Philosophical Transactions of the {R}oyal Society of
  {L}ondon}, \textbf{53} 370–418.

\bibitem[{Bellhouse(2011)}]{bellhouse:2011}
\textsc{Bellhouse, D.} (2011).
\newblock \textit{Abraham {D}e {M}oivre}.
\newblock CRC Press, Boca Raton.

\bibitem[{Dale(1982)}]{dale:1982}
\textsc{Dale, A.~I.} (1982).
\newblock Bayes or {L}aplace? An examination of the origin and early
  application of {B}ayes' theorem.
\newblock \textit{Archive for the History of the Exact Sciences}, \textbf{27}
  23–47.

\bibitem[{Dale(1999)}]{dale:1999}
\textsc{Dale, A.~I.} (1999).
\newblock \textit{A History of Inverse Probability}.
\newblock Springer-Verlag, New York.
\newblock (Second edition.).

\bibitem[{Keynes(1920)}]{keynes:1921}
\textsc{Keynes, J.} (1920).
\newblock \textit{A {T}reatise on {P}robability}.
\newblock Macmillan and Co., London.

\bibitem[{Laplace(1812)}]{laplace:1812}
\textsc{Laplace, P.} (1812).
\newblock \textit{Th\'eorie Analytique des Probabilit\'es}.
\newblock Courcier, Paris.

\bibitem[{McGrayne(2011)}]{mcgrayne:2011}
\textsc{McGrayne, S.} (2011).
\newblock \textit{The Theory that Would Not Die}.
\newblock Yale Univ Press, New Haven, CT.

\end{thebibliography}
\end{document}